\documentclass[aps,prb,twocolumn,superscriptaddress,showpacs]{revtex4-1}
\usepackage{amsmath,amssymb,amsfonts,amsthm}
\usepackage{graphicx}
\usepackage{bm}
\usepackage{bbm}
\usepackage{color}
\usepackage{dcolumn}   
\usepackage{epstopdf}
\usepackage[caption=false]{subfig}
\usepackage{xr}
\usepackage[colorlinks=true,linkcolor=blue,urlcolor=blue,citecolor=blue]{hyperref}

\begin{document}

 \newcommand{\breite}{1.0} 

\newtheorem{prop}{Proposition}
\newtheorem{cor}{Corollary} 

\newcommand{\be}{\begin{equation}}
\newcommand{\ee}{\end{equation}}

\newcommand{\bea}{\begin{eqnarray}}
\newcommand{\eea}{\end{eqnarray}}
\newcommand{\lt}{<}
\newcommand{\gt}{>} 

\newcommand{\Reals}{\mathbb{R}}     
\newcommand{\Com}{\mathbb{C}}       
\newcommand{\Nat}{\mathbb{N}}       

\newcommand{\id}{\mathbboldsymbol{1}}    

\newcommand{\Real}{\mathop{\mathrm{Re}}}
\newcommand{\Imag}{\mathop{\mathrm{Im}}}

\def\O{\mbox{$\mathcal{O}$}}   
\def\F{\mathcal{F}}			
\def\sgn{\text{sgn}}

\newcommand{\deo}{\ensuremath{\Delta_0}}
\newcommand{\dea}{\ensuremath{\Delta}}
\newcommand{\ak}{\ensuremath{a_k}}
\newcommand{\ad}{\ensuremath{a^{\dagger}_{-k}}}
\newcommand{\sx}{\ensuremath{\sigma_x}}
\newcommand{\sz}{\ensuremath{\sigma_z}}
\newcommand{\spl}{\ensuremath{\sigma_{+}}}
\newcommand{\smi}{\ensuremath{\sigma_{-}}}
\newcommand{\alk}{\ensuremath{\alpha_{k}}}
\newcommand{\bk}{\ensuremath{\beta_{k}}}
\newcommand{\ok}{\ensuremath{\omega_{k}}}
\newcommand{\vd}{\ensuremath{V^{\dagger}_1}}
\newcommand{\vi}{\ensuremath{V_1}}
\newcommand{\vo}{\ensuremath{V_o}}
\newcommand{\zc}{\ensuremath{\frac{E_z}{E}}}
\newcommand{\xc}{\ensuremath{\frac{\Delta}{E}}}
\newcommand{\xd}{\ensuremath{X^{\dagger}}}
\newcommand{\aok}{\ensuremath{\frac{\alk}{\ok}}}
\newcommand{\tpw}{\ensuremath{e^{i \ok s }}}
\newcommand{\tpe}{\ensuremath{e^{2iE s }}}
\newcommand{\tmw}{\ensuremath{e^{-i \ok s }}}
\newcommand{\tme}{\ensuremath{e^{-2iE s }}}
\newcommand{\epls}{\ensuremath{e^{F(s)}}}
\newcommand{\emis}{\ensuremath{e^{-F(s)}}}
\newcommand{\epl}{\ensuremath{e^{F(0)}}}
\newcommand{\emi}{\ensuremath{e^{F(0)}}}

\newcommand{\lr}[1]{\left( #1 \right)}
\newcommand{\lrs}[1]{\left( #1 \right)^2}
\newcommand{\lrb}[1]{\left< #1\right>}
\newcommand{\nbt}{\ensuremath{\lr{ \lr{n_k + 1} \tmw + n_k \tpw  }}}

\newcommand{\om}{\ensuremath{\omega}}
\newcommand{\dw}{\ensuremath{\Delta_0}}
\newcommand{\wbp}{\ensuremath{\omega_0}}
\newcommand{\dv}{\ensuremath{\Delta_0}}
\newcommand{\vbp}{\ensuremath{\nu_0}}
\newcommand{\vplus}{\ensuremath{\nu_{+}}}
\newcommand{\vminus}{\ensuremath{\nu_{-}}}
\newcommand{\wplus}{\ensuremath{\omega_{+}}}
\newcommand{\wminus}{\ensuremath{\omega_{-}}}
\newcommand{\uv}[1]{\ensuremath{\mathbf{\hat{#1}}}} 
\newcommand{\abs}[1]{\left| #1 \right|} 
\newcommand{\norm}[1]{\left \lVert #1 \right \rVert} 
\newcommand{\avg}[1]{\left< #1 \right>} 
\let\underdot=\d 
\renewcommand{\d}[2]{\frac{d #1}{d #2}} 
\newcommand{\dd}[2]{\frac{d^2 #1}{d #2^2}} 
\newcommand{\pd}[2]{\frac{\partial #1}{\partial #2}} 
\newcommand{\pdd}[2]{\frac{\partial^2 #1}{\partial #2^2}} 
\newcommand{\pdc}[3]{\left( \frac{\partial #1}{\partial #2}
 \right)_{#3}} 
\newcommand{\ket}[1]{\left| #1 \right>} 
\newcommand{\bra}[1]{\left< #1 \right|} 
\newcommand{\braket}[2]{\left< #1 \vphantom{#2} \right|
 \left. #2 \vphantom{#1} \right>} 
\newcommand{\matrixel}[3]{\left< #1 \vphantom{#2#3} \right|
 #2 \left| #3 \vphantom{#1#2} \right>} 
\newcommand{\grad}[1]{{\nabla} {#1}} 
\let\divsymb=\div 
\renewcommand{\div}[1]{{\nabla} \cdot \boldsymbol{#1}} 
\newcommand{\curl}[1]{{\nabla} \times \boldsymbol{#1}} 
\newcommand{\laplace}[1]{\nabla^2 \boldsymbol{#1}}
\newcommand{\vs}[1]{\boldsymbol{#1}}
\let\baraccent=\= 

\def\red#1{{\textcolor{red}{#1}}}

\title{Dynamical enhancement of symmetries in many-body systems}

\author{Kartiek Agarwal}
\email{agarwal@physics.mcgill.ca}
\affiliation{Department of Physics, McGill University, Montr\'{e}al, Qu\'{e}bec H3A 2T8, Canada}
\author{Ivar Martin}
\affiliation{Material Science Division, Argonne National Laboratory, Argonne, IL 08540, USA}

\date{\today}
\begin{abstract}

We construct a dynamical decoupling protocol for accurately generating local and global symmetries in general many-body systems. Multiple commuting and non-commuting symmetries can be created by means of a self-similar-in-time (``polyfractal") drive. The result is an effective Floquet Hamiltonian that remains local and avoids heating over exponentially long times. This approach can be used to realize a wide variety of quantum models, and non-equilibrium quantum phases. 
\end{abstract}
\maketitle

\paragraph*{\textbf{Introduction.---}}

 Much of the richness of the material universe transpires due to a sequence of spontaneous symmetry breaking events, going from the highest (most symmetric) energy scales down to the lowest. By definition, every equilibrium physical system has already found its energetic low symmetry optimum; can it nevertheless be re-purposed to realize a different spontaneous symmetry breaking pathway? 
 
 Sometimes this can be achieved by tuning thermodynamic parameters. For instance, hydrostatic pressure can restore more symmetric crystalline phases at a given temperature, or magnetic fields can suppress superconductivity to reveal other competing instabilities. Such thermodynamic knobs are, unfortunately, quite limited. 

A more flexible approach is to re-institute symmetries dynamically. In nuclear magnetic resonance, echo techniques have long been used to improve the coherence of local moments by dynamically suppressing their coupling to the environment. The Hahn echo\cite{hahn} for instance reduces inhomogeneous broadening, and the WAHUHA protocol\cite{wahuha} can suppress anistropic dipole-dipole interactions. However, the extension of these ideas to creating global symmetries in many-body, interacting systems is less clear. The main concern is that driving concomitantly generates heating in these systems, which in turn suppresses interesting collective phenomena. 

In this Letter, we discuss dynamical protocols that can be used to engineer multiple global or local symmetries, while keeping heating at bay for exponentially long times.  This  paves the way to creating novel symmetry broken and topological phases~\cite{senthil2015symmetry,wen2017zoo} out of low symmetry templates. 

%
%
%
Our work is informed by recent progress in understanding Floquet dynamics of many-body quantum systems. Crucially, it has been shown that the naive expectation that driving should inevitably lead to heating is not always correct---strong disorder~\cite{mbldrivingnonheatingLazarides2015,khemanipispinglass,pontedrivenMBL} and/or appropriate drive frequency selection~\cite{ADHH,abanin2016theory,MoriSaitoKuwahara,kuwahara2016floquet,elseprethermal} can push heating to exponentially long times. Further, in the context of time-crystals, it is already well appreciated that driving may lead to the creation of $\mathcal{Z}_2$ symmetry whose spontaneous rupturing gives rise to time-crystalline phase~\cite{keyserlingkAbelian,elseprethermal,elsefloquet}. In this work, we show how such ideas can be extended to generate multiple local and global symmetries in the effective Floquet Hamiltonian. 

We demonstrate our protocol by considering the specific case of how multiple $\mathcal{Z}_2$ symmetries can be generated in spin systems. (Generalization to the $\mathcal{Z}_n$ case with $n>2$ is straightforward.) The protocol involves injecting a finite set, of say, $n_s$ unitary operators $X_i$, at specific times corresponding to a fractal pattern, in between regular unitary evolution under the system's physical Hamiltonian $H$. As we show, such fractal application of $X_i$ can be optimized in the number of fractal layers $n_f$, to result in an effective Hamiltonian for which $X_i$s are symmetries to an accuracy that is nearly exponential in the drive frequency. This sensitivity to the drive frequency allows for accurately implementing \emph{global} symmetries while requiring a drive frequency that scales merely logarithmically with system size $N$. 

\begin{center}
\begin{figure}
\includegraphics[width=3.3in]{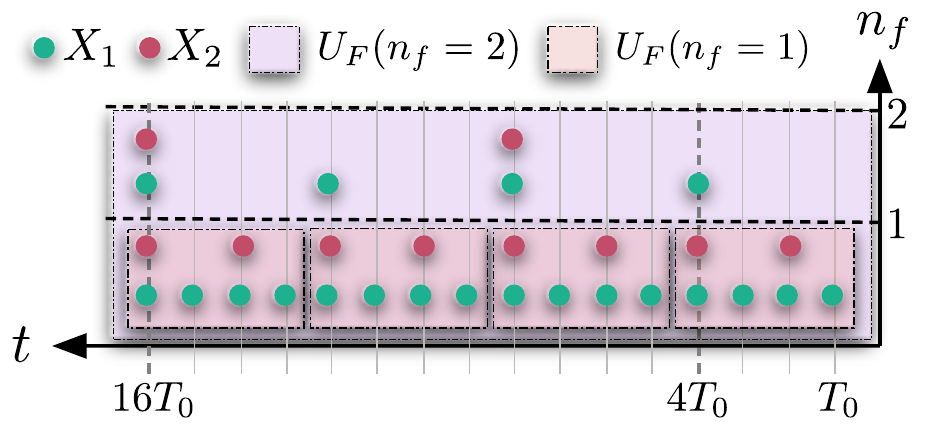}
\caption{Illustration of the protocol for $n_s = 2$ unitaries $X_1, X_2$. The two unitaries are applied in a self-similar fashion in time. The Floquet unitary $U_F (T \equiv 2^{n_f n_s} T_0)$ at fractal layer $n_f$, is the product of operators applied at the times indicated, from bottom up. Note also that $X^2_i = 1$, and $X_i$s either commute or anti-commute with one another. Thus, the net operator applied at any time step is either $X_1,X_2,X_1 X_2$, or the identity.}
\label{fig:figprotocol}
\end{figure}
\end{center}
\vspace{-0.85cm}

The approach is motivated by the following intuition. The periodic application of $X_i$ flips the sign of terms in $H$ that anti-commute with $X_i$;  thus, frequent application of $X_i$ leads to their cancellation after even number of drive periods. The resulting effective Hamiltonian commutes with $X_i$ up to $\sim \mathcal{O}(T_0)$ corrections, where $T_0$ is the drive period. As we show, these corrections can be further suppressed by applying $X_i$ periodically at intervals of $2T_0, 4T_0, ...$ (see Fig.~\ref{fig:figprotocol}). (Note since $X^2_i = 1$, $X_i$ will effectively not be applied at some times.)  Alternatively, one may apply other generators $X_j$s at self-similar intervals to generate \emph{additional} symmetries. This work follows through with the above intuition, and shows that for local Hamiltonians, there exists an optimal number of fractal layers $n_f$ which leads to superpolynomial (in drive-frequency) suppression of symmetry-violating terms, while heating occurs on a stretched-exponentially long timescale. 

This work has implications for engineering novel Floquet Hamiltonians and non-equilibrium phases~\cite{moessner2017equilibration,kitagawatopologicaldriven,Lindnerchiraldriven,rudnertwodimensionaldriventopo,keyserlingkAbelian,baireydrivinginducedmbl,choi2018dynamically,yao2017discrete,AgarwalStronglyDrivenAnderson}, symmetry-protected topological phases~\cite{senthilSPT}, and stabilization of quantum memories~\cite{lidarhamiltoniancancel,faulttolerantDDLidar}, among others. When the engineered symmetries anti-commute, they give rise to a robust degeneracy structure in the entire spectrum. This potentially could be exploited to engineer topologically protected qubits manipulated by the same $X_i$ to high precision. We explore some of these ideas in a companion paper~\cite{martin2020textit}.These ideas should also easily translate to experiments---particularly in setups exploring non-equilibrium quantum phases, such as trapped ions~\cite{zhang2017observation}, NV centers~\cite{choi2017observation}, among others~\cite{bordia2017periodically,bernien2017probing}---by introducing multiple drives, akin to those already used, but operating at multiples of the base frequency.    

Finally, note that there is a precedent for fractal pulsing in finite-sized systems~\cite{lidarhamiltoniancancel,faulttolerantDDLidar}. However, as opposed to the more general scheme we identify, these are system specific, and rely on operator expansions that have null convergence in the thermodynamic limit~\cite{ADHH}. One may thus view this work as a formal extension of dynamical decoupling techniques to many-body systems. Below we describe our results, before providing detailed proofs and numerical validation. 

\paragraph*{\textbf{Description of protocol and main results.---}}

We assume a physical system described by a Hamiltonian $H$ comprising of a sum of local terms, with a local norm $\norm{h}$. We also assume a set of $n_s$ unitaries $X_i$ that further satisfy the condition $X^2_i = 1$. These unitaries may themselves either commute or anti-commute with one another. The protocol we study involves applying $X_i$ at times 
\begin{align}
t_i = m 2^{i + n_s (j-1)} T_0, \; \; i \in [1, n_s], j \in [1,n_f], m \in \mathcal{Z}^+
\end{align}
amidst the regular Heisenberg evolution, where $n_f$ is the number of ``fractal layers'' in the composite Floquet unitary; see Fig.~\ref{fig:figprotocol} for illustration. To illustrate by example, suppose we have $n_s = 2$ unitaries, and apply these $n_f = 1$ times. Then, $U(T_0) = e^{- i H T_0}$; $U(2 T_0) = X_1 U(T_0) X_1 U(T_0)$; $U (4 T_0) = X_2 U (2T_0) X_2 U(2 T_0)$. Subsequent time-evolution at periods of $T \equiv 2^{n_f n_s} T_0 = 4 T_0$ is given by the repeated application of the Floquet unitary $U(4T_0)$. For $n_f = 2$, the above recursion relations would be repeated for another fractal layer: $U(8 T_0) = X_1 U(4 T_0) X_1 U(4T_0)$; $U(16 T_0) = X_2 U(8 T_0) X_2 U (8 T_0)$, and $U(16 T_0)$ would subsequently serve as the Floquet unitary. 

We now decompose $H$ into terms which transform differently under $X_i$:  
\begin{align}
&H = \sum_{\vs{\epsilon}} A_{\vs{\epsilon}} \; \text{where} \; \vs{\epsilon} = (\epsilon_1, ..., \epsilon_{n_s}), \epsilon_i \in \{ 0,1 \}, \nonumber \\
&X_j A_{\vs{\epsilon}} X_j = (-1)^{\epsilon_j} A_{\vs{\epsilon}}. 
\end{align}
This decomposition is unique if $X_i$s commute or anti-commute with one another, which we assume. 
With this terminology, one may represent the Floquet unitary in time-ordered notation as 
\begin{align}
&U (T \equiv 2^{n_f n_s} T_0) = \mathcal{T} \left\{ e^{-i \int^T_0 dt \; \sum_{\vs{\epsilon}} A_{\vs{\epsilon}} f_{\vs{\epsilon}} (t) } \right\}, \nonumber \\
&\text{where} \; f_{\vs{\epsilon}} (t) = \pm 1 \;  \; \text{and} \; \int_0^{2^{n_s n_f}T_0} f_{\vs{\epsilon}} (t) = \delta_{\vs{0}, \vs{\epsilon}}. \label{eq:U}
\end{align}

Here $f_\epsilon(t)$ tracks times at which $X_i$ is applied; this corresponds to a sign change of terms $A_{\vs{\epsilon}}$ for which $\epsilon_i = 1$. The integral over a complete period is zero except for $f_{\vs{0}}$. Thus, in a time-averaged sense, the effective Hamiltonian is $A_{\vs{0}}$ comprising of only terms even under all $X_i$. 

Now, one may represent unitary $U(T)$ as an expansion in the exponent  
\begin{align}
U(T) = e^{- i T \sum_{n=0}^{\infty} T^n \Omega_n}
\label{eq:expansion}
\end{align} 
with operators $\Omega_n$ that can be arrived at using the Magnus expansion, or, in this case, a repeated application of the BCH formula; the first term is simply the time-averaged Hamiltonian $\Omega_0 = A_{\vs{0}}$. 

In general, the operator $\Omega_n$ involves $n$ nested commutators of the local operators $A_{\vs{\epsilon}}$. Thus, if the local terms comprising $A_{\vs{\epsilon}}$ involve at most $k$ sites, $\Omega_n$ can be represented as a sum of terms comprising at most $nk$ sites. Finally, the series expansion is only useful if we can truncate it at some order and effectively approximate the unitary dynamics; we define the approximate Hamiltonian
\begin{align}
H^{(n_0)}_F = \sum_{0\le n \le n_0} T^n \Omega_n. 
\label{eq:Hexpansion}
\end{align}
Our main results concern the properties of the Floquet unitary $U(T)$, and the associated effective Floquet Hamiltonian $H^{(n_0)}_F$. The first part of our results are directly adapted from the results of Refs.~\cite{MoriSaitoKuwahara,kuwahara2016floquet,ADHH,abanin2017rigorous}, which state that the difference between the exact reduced density matrix of a region of size $N_\rho$ and that obtained by evolving it with $H^{(n_0)}_F$ is bounded in norm by $c N_\rho 2^{-n_0} $, for some finite constant $c$. Here $n_0 \sim 1 / (T \norm{h}) \equiv \omega / \norm{h} \gg 1$ scales linearly with the effective drive frequency $\omega = 1/ \left( 2^{n_f n_s} T_0 \right)$. The norm of this error sets the inverse of the time scale up to which $H^{(n_0)}_F$ provides a good description of the dynamics of local operators (alternatively, the time scale for heating)---crucially, this timescale grows exponentially with the drive frequency. For global operators $X_i$ with finite norm, but also for $H^{(n_0)}_F$ itself, $N_\rho$ is the system size $N$, but the error can still be made small by scaling $\omega$ merely logarithmically with system size. Hereon, we will assume such frequency scaling. 

Having established the conditions under which $H^{(n_0)}_F$ faithfully describes the time evolution of $X_i$, we seek to establish a bound on the norm of terms in $H^{(n_0)}_F$ that do not commute with $X_i$. Defining the time scale $\tau_X \approx \text{min}_i \left \{ 1/\norm{\left[ H^{(n_0)}_F , X_i \right]} \right\}$ which sets the shortest timescale at which unitaries $X_i$ relax, we find 
\begin{align}
\tau_X &\ge \frac{1}{N} (c_1 2^{n_s n_f} T_0 \norm{h} n_f )^{-n_f}, \nonumber \\
\tau_H &\ge \frac{1}{N} e^{c_2 \frac{1}{T_0 \norm{h}} \cdot \frac{1}{2^{n_s n_f}}}. 
\label{eq:eqXH1}
\end{align} 
where $c_1,c_2$ are $\mathcal{O}\left( 1 \right)$ combinatorial constants.  $\tau_H$ is a bound on the time for which global operators such as energy are accurately described by $H^{(n_0)}_F$ \cite{MoriSaitoKuwahara,ADHH}. Note that $\tau_X$ initially increases with the number of fractal layers $n_f$ but eventually begins to decrease again. There is therefore an optimal $n_f$ for which $X_i$ become effective symmetries. Note also that $n_f$ cannot be made arbitrary large since its increase rapidly decreases the thermalization time scale $\tau_H$.

We now describe how to optimize $n_f$ to maximize $\tau_X, \tau_H$. First, note that to maintain exponential dependence on the reference 
drive frequency $\omega_0 \equiv 1/T_0$, $n_f$ must scale at most logarithmically in the small parameter $T_0 \norm{h}$. This implies $n_f = \frac{x}{n_s} \text{log}_2 \left( \frac{1}{T_0 \norm{h}} \right)$ with $0 < x < 1$. Plugging this into the result for $\tau_X$, we find 
\begin{align}
\tau_X &\ge \frac{1}{N} \left(c'_2 \cdot \frac{\abs{\text{log}_2 \left( T_0 \norm{h} \right)}}{ T_0 \norm{h}}\right)^{\left( 1- x \right) \frac{x}{n_s} \text{log}_2 \left( \frac{1}{T_0 \norm{h}} \right)}, \nonumber \\
\tau_H &\ge \frac{1}{N} e^{c_2 \frac{1}{(T_0 \norm{h})^{1-x}}}, \; \; \text{for some} \; 0 < x < 1. 
\label{eq:eqXHfin}
\end{align} 
where $c'_2$ is an $\mathcal{O}(1)$ constant. Thus, we can vary $\omega_0$ to control $\tau_H$ with (stretched-) exponential sensitivity, and $\tau_X$ as a power-law that can be made arbitrarily large. Consequently, a very slow increase of $\omega_0$ with system size $N$ is sufficient to cancel the prefactor of $1/N$ in both $\tau_X$ and $\tau_H$. 

Finally, we note that $H^{(n_0)}_F$ is quasi-local in the sense that the amplitude of terms decays exponentially with the spatial range~\cite{MoriSaitoKuwahara,ADHH}. In general, this operator may be hard to evaluate exactly, but it can be approximated by
\begin{align} 
H^{(n_0)}_F \approx H^{(0)}_F = A_{\vs{0}}\label{eq:A}
\end{align}

where $A_{\vs{0}}$ commutes with all $X_i$ by construction. Since it captures the time-evolution of local operators and, importantly, also reflects the global symmetry properties of $H^{(n_0)}_F$, it is a good approximation to the effective Floquet Hamiltonian for times $t \lesssim \text{min} \left( \tau_X, \tau_H \right)$.

 The design of the protocol, which is crucial to the bound obtained in the first part of Eq.~(\ref{eq:eqXH1}), its proof, and the result of Eqs.~(\ref{eq:eqXHfin}) that times $\tau_X$ and $\tau_H$ are almost exponentially sensitive to the base drive frequency are the central results of this work. 

\paragraph*{\textbf{Fractal driving with a single unitary $X$.---}}
We now derive the bound in the first part of Eqs.~(\ref{eq:eqXH1}) for the case ($n_s = 1$) of a single unitary $X_1 \equiv X$. The derivation of the result will also help the reader intuit the logic behind fractal driving. 

Using the terminology introduced above, the Hamiltonian is composed of two (kinds of) terms: $H = A_0 + A_1$, where $A_{0 (1)}$ is even (odd) under $X$. In this case, fractal driving can be described by the simple recurrence relations
\begin{align}
U(2^n T_0) = X \cdot U (2^{n-1} T_0) \cdot X \cdot U ( 2^{n-1} T_0), \forall n \ge 1 
\label{eq:recur}
\end{align}
with $ U(T_0) = e^{-i H T_0}$. At the first stage, this implies
\begin{align}
U(2 T_0) = e^{-i T_0 ( A_0 - A_1 ) } e^{- i T_0 ( A_0 + A_1)} \equiv e^{- i T^{(1)} (A^{(1)}_0 + A^{(1)}_1)},  
\end{align}
where we define $T^{(1)} \equiv 2 T_0$, and $A^{(1)}_0$ and $A^{(1)}_1$ are the new effective terms that are even and odd, respectively, under $X$. The BCH formula then yields
\begin{align}
A^{(1)}_0 &= A^{}_0 + \mathcal{O} \left( T^2_0 \right), \nonumber \\
A^{(1)}_1 &= -i \frac{T_0}{2} \left[ A_0^{}, A_1^{} \right] + \mathcal{O} \left( T^2_0 \right).
\label{eq:recur1}
\end{align}
After $n_f$ fractal layers, this implies
\begin{align}
T^{(n_f)} = & 2^{n_f}T_0, \; \; \; A^{(n_f)}_0 = A_0, \nonumber \\
A^{(n_f)}_1 = & \left(-i 2^{\frac{n_f -3}{2}} T_0 \right)^{n_f}  \underbrace{[ A_0, ... , [ A_0}_{n_f},  A_1 ]...] + \mathcal{O} \left( T^{n_f + 1}_0 \right), \nonumber \\
\label{eq:normest}
\end{align}

Importantly, terms in $H^{(n_0)}_F$ that anti-commute with $X$ appear first at $\mathcal{O}\left(T_0^{n_f} \right)$. These terms are a subset of all terms that appear at $\mathcal{O} \left( T^{n_f} \right)$ in the expansion of the Floquet Hamiltonian, Eq.~(\ref{eq:Hexpansion}).  Their norm is therefore bounded by $T^{n_f} \norm{\Omega_{n_f}}$. Further, the norm of \emph{all} terms that may anti-commute with $X$ can be bounded by $\sum_{n=n_f}^{n_0} T^n \norm{\Omega_{n}}$. We note from Ref.~\cite{MoriSaitoKuwahara}, that 
\begin{align}
    \norm{\Omega_n} T^n \le N \frac{(c T \norm{h})^n n!}{(n+1)^2} \le N (c T \norm{h} n)^n
    \label{eq:moribound}
\end{align}
for some $\mathcal{O} (1)$ constant $c$. Using the above, we can bound the ratio $\norm{\Omega_{n+1}}/\norm{\Omega_{n}} < 1/2 \; \forall n \le n_0$, if we set $n_0 = 1/(2 c T \norm{h})$. This finally implies 
\begin{align}
    \norm{\left[H^{(n_0)}_F,X\right]} \le 2 T^{n_f} \norm{\Omega_{n_f}}. 
\label{eq:fin1}
\end{align}
Eqs.~(\ref{eq:fin1}) and~(\ref{eq:moribound}) give the result in Eq.~(\ref{eq:eqXH1}) for $n_s = 1$. 
\begin{center}
\begin{figure*}[ht]
\includegraphics[width=\textwidth]{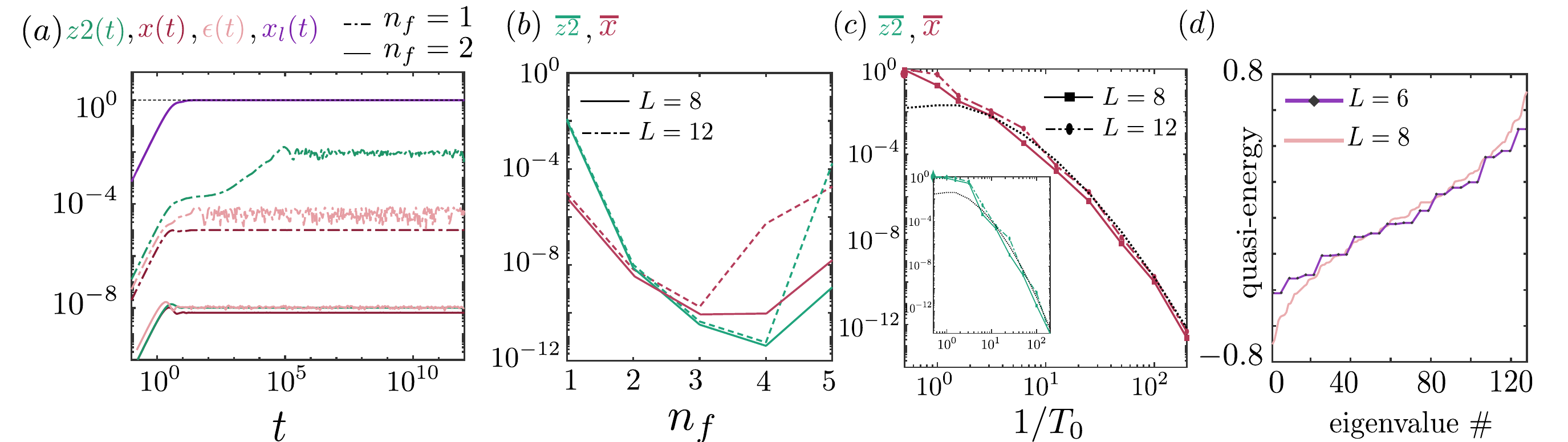}
\caption{(a) The time-dependent relaxation of various global and local operators (see main text), and heating is plotted in a system driven by $P_X, P_{Z2}$. Initial relaxation on microscopic timescales gives way to a long prethermal regime. (b) The long time values $\bar{x}, \bar{z2}$ of operators $P_X, P_{Z2}$ depend in a non-monotonic way on fractal layers $n_f$, with optimal $n_f = 3$ in this instance. For (a),(b), $T_0 = 0.01$.  (c) The long-time relaxation is strongly suppressed with increasing frequency $1/T_0$. Black dotted lines are theoretical curves of the form $f (T_0) \sim \left(\frac{aT_0}{log_2(aT_0)}\right)^{b \text{log}_2 (aT_0)}$ [for $a = 1, b = 0.43 \; (0.5 \; \text{for inset})$], as for $1/\tau_X$ in Eq.~(\ref{eq:eqXHfin}). (d) $P_X, P_{Z2}$ (anti-) commute for $L = (6) 8$ yielding a doubly degenerate spectrum for $L = 6$ but not for $L = 8$.}
\label{fig:numerical}
\end{figure*}
\end{center}

\paragraph*{\textbf{Polyfractal driving with multiple unitaries.---}}
We now generalize to the case $n_s > 1$. As before, we can examine the flow of $A_{\vs{\epsilon}}$ after each fractal layer, that is, at times $2^{n n_s} T_0$ for integer $n$. (Recall, a fractal layer corresponds to the application of each $X_i$ once at progressively doubled periods.)

Let us examine the recursion relations for $n_s = 2$ to illustrate by example.
The Hamiltonian in this case is $H = A_{00} + A_{01} + A_{10} + A_{11}$, where terms $A_{01}$ and $A_{10}$ are odd under $X_1$ and $X_2$ only, respectively, while $A_{11}$ is odd under both. $U(4T_0)$ is calculated using the BCH expansion twice. Representing it as $U(4 T_0) = e^{- i T^{(1)} \left( A^{(1)}_{00} + A^{(1)}_{01} + A^{(1)}_{10} + A^{(1)}_{11} \right) }$, where $T^{(1)} = 4 T_0$, we find to $\mathcal{O} \left( T^2_0 \right)$

\begin{align}
&A^{(1)}_{00} = A_{00}, \; A^{(1)}_{01} = -i T_0 [ A_{00} , A_{01} ], \nonumber \\
&A^{(1)}_{10} = - i \frac{T_0}{2} \left( [A_{00} , A_{10}] + [A_{01}, A_{11}] \right), \nonumber \\
&A^{(1)}_{11} = - i T_0 \left( -i \frac{T_0}{2} \right) \left[ A_{00}, [A_{00}, A_{11}] + [A_{01}, A_{10}] \right] \nonumber \\
&+ i T_0 \left( -i \frac{T_0}{2} \right)  \left[ A_{01}, [A_{00}, A_{10}] + [A_{01}, A_{11}] \right] . 
\end{align}

Note that terms which are odd under just one symmetry are canceled to $\mathcal{O} \left( T_0 \right)$, while terms odd under both $X_1,X_2$ are canceled to higher order. Similar conclusions apply for the general case of $n_s \ge 2$. 

After $n_f$ fractal layers, symmetry-violating terms appear at order $\mathcal{O}\left(T^{n_f}_0\right)$ or higher. (The terms that are odd under just one symmetry appear at the lowest order.) To estimate the norm of these terms, we can apply the same arguments for the case $n_s = 1$, arriving at the results of Eqs.~(\ref{eq:eqXH1}). This completes the proof. 

\paragraph*{\textbf{Numerical Results.---}}

We now provide numerical simulations to illustrate the above results. We consider a short-ranged spin-$1/2$ chain of length $L$, with open boundary conditions. In the majorana representation, the Hamiltonian reads
\begin{align}
H = \sum_{n,k \le 4} e^{-k+1} \left( -i \gamma_n \gamma_{n+k} \right) + V \gamma_n \gamma_{n+1} \gamma_{n+2} \gamma_{n+3}
%
\end{align}
where for $n$ odd/even, $\gamma_n = \prod_{j<n} \sigma^z_j \sigma^{x/y}_n$. $H$ has parity symmetry $P_Z = \prod_n \sqrt{i} \gamma_n$; we work in the sector $P_Z = 1$. 

One can check that driving with $P_X = \prod_j \sigma^x_j = \prod_n i \gamma_{4n-2}\gamma_{4n-1}$ suppresses \emph{even} nearest neighbor  majorana bonds and yields a Kitaev chain with terminal majorana zero modes. 
However, the quality (energy splitting) of the majoranas {\em is not} equivalent to the accuracy of $P_X$ symmetry; for details see Supplemental Material (SM). 

To illustrate the effectiveness of the protocol in creating multiple global symmetries, we  drive the system with $P_X$ and $P_{Z2} = \prod_j \sigma^{z}_{2 j} = \prod_{n} i \gamma_{4n-3} \gamma_{4n - 2}$. These operators (anti-) commute for $L = (4n+2) \; 4n$ for integer $n$. For $L = 4n+2$, the operators satisfy the Pauli algebra which leads to a doubly-degenerate spectrum.

To quantify the accuracy of the generated symmetries, we evaluate ``decoherences"--- $a(t) = 1 - \text{Tr}\left[ P_{a} (t) P_{a} \right] / 2^{L-1}$ of relevant operators $P_a$. $a(t = 0) = 0$ and remains zero for perfectly conserved $P_a$, while it relaxes to $1$ for non-conserved operators. We study $P_a = \{ P_X, P_{Z2}, \sigma^x_{L/2} \sigma^x_{L/2+1} \}$. 
The first two measure the conservation of $P_X, P_{Z2}$, and should yield $x(t) = z2(t) = 0$ in case they are perfect symmetry generators of the Floquet dynamics, while $x_l (t)$ measures the relaxation of a local operator that is not expected to be conserved. Finally, we compute $\epsilon (t) = \avg{A_{\vs{0}} (t)-A_{\vs{0}} (0)}$  where the average is taken with respect to the  ground state of $A_{\vs{0}}$ [the part of $H$ that commutes with  $P_X$ and $P_{Z2}$, see Eq. (\ref{eq:A})]. $\epsilon (t)$ thus characterizes heating in the system.  

The numerical results of Fig.~\ref{fig:numerical} (a) generically exhibit rapid initial relaxation on microscopic timescales, before transitioning to a long-lived prethermal state. This is seen via the initial rapid loss of coherence of $P_{Z2}$, $P_X$ and increase in the energy $\epsilon (t)$, before plateauing at a fixed value much smaller than $1$. The decoherence $x_l (t)$ on the other hand rapidly approaches $1$, as expected. Long-term values of the decoherences, $\overline{z2}, \overline{x}$ are seen to improve as fractal layers are increased from $n_f = 1$, degrading subsequently for larger $n_f$---see Fig.~\ref{fig:numerical} (b)---illustrating the existence of an optimal number of fractal layers for symmetry creation. Fig.~\ref{fig:numerical} (c) illustrates the sensitivity of long-time coherences in the optimal protocol (over $n_f$) to $T_0$, as expected. Finally, in Fig.~\ref{fig:numerical} (d), we confirm our expectations that the eigenspetrum of the Floquet unitary is doubly degenerate for $L = 6$ and not for $L = 8$. 

An interesting aspect of the numerical results is that unlike our expectations, we do not observe heating away from the ``prethermal plateau'' at times longer than $\tau_X$ (Eq.~(\ref{eq:eqXHfin}). 
However, the plateau values of  $\overline{x}, \overline{z2}$ appear to scale with $1/\tau_X$. In fact, we observe relaxation (to $1$) only when driving at frequencies smaller than the microscopic scale, or for local operators not designed to be conserved. 
Whether this is a limitation of  the small system sizes, a feature of the particular model that we considered, or an indication that our protocol works generally better than the conservative estimate for heating that we made, deserves further study.


\paragraph*{\textbf{Summary and Outlook.---}}
We have introduced a novel strong-driving protocol for engineering Floquet Hamiltonians,  by creating new local and global symmetries.  It may be viewed as an extension of dynamical decoupling techniques to local many-body Hamiltonians. While we describe here creation of $\mathcal{Z}_2$ symmetry generators, the results are easily generalized to $Z_{n>2}$ by applying the individual symmetry generators in sets of $n$ instead of twice, as in Eq. (\ref{eq:recur}). 

The symmetries can be used to engender a variety of novel Hamiltonians and dynamical phenomena. Creation of topological phases, and quantum memory stabilization using such schemes are explored in Ref.~\cite{martin2020textit}. While we have focused on the quasi-stationary Hamiltonian $H_F$, the dynamics of the system inside the Floqut period can be subject to interesting dynamical phenomena and deserves further attention.

Experiments probing non-equilibrium phenomena in driven systems in a variety of setups including Nitrogen-vacancy centers~\cite{choi2017observation}, ion traps~\cite{zhang2017observation}, cold atoms~\cite{bordia2017periodically,bernien2017probing} among others would be the natural setup to explore these ideas. 

\paragraph*{\textbf{Acknolwedgements.---}} We thank Jonathan Baugh for pointing out the ``concatenated driving'' protocol proposed in Refs.~\cite{faulttolerantDDLidar,lidarhamiltoniancancel,khodjastehlidarviolaPRL1,cai2012robust} for protecting information in single qubits, which bears resemblance to our approach, and Lorenza Viola for pointing out previous work where dynamical decoupling schemes have been used for Hamiltonian simulation~\cite{santos2008advantages,bookatz2014hamiltonian}. We also thank B. Bauer, W. A. Coish, A. Pal, T. Pereg-Barnea, D. Pikulin, and L. Viola for valuable discussions. KA acknowledges support from NSERC Grants RGPIN-2019-06465, and DGECR-2019-00011,  and start-up funds from McGill University for support. Work at Argonne was supported by the Department of Energy, Office of Science, Materials Science and Engineering Division.

\bibliographystyle{apsrev4-1}
\bibliography{polyfrac}

\newpage 

\section{Kitaev Chain Example}

We study the model as introduced in the main text, with $V = e^{-3}$. We drive the system with the unitary $P_X = \prod_i \sigma^i_x$. In the majorana formalism, this corresponds to the operator 

\be
P_X = \gamma_{2} \gamma_{3} \gamma_{6} \gamma_{7} ... = \prod_i \gamma_{4i-2} \gamma_{4i-1}
\ee

One can easily confirm that the Hamiltonian $H_{0} = (H + P_X H P_X)/2$ which is invariant under $P_X$ exactly has weak and strong alternating bonds between nearest-neighbor majorana fermions. It is important to note that an imperfect weakening of the odd set of bonds is \emph{sufficient} to drive the majorana system into the Kitaev phase (provided interactions are weak enough). In other words, provided the $\mathcal{Z}_2$ fermion parity is perfectly conserved, one does not need to introduce additional symmetries to enter the Kitaev phase. 

\begin{center}
\begin{figure}[ht]
\includegraphics[width=3.5in]{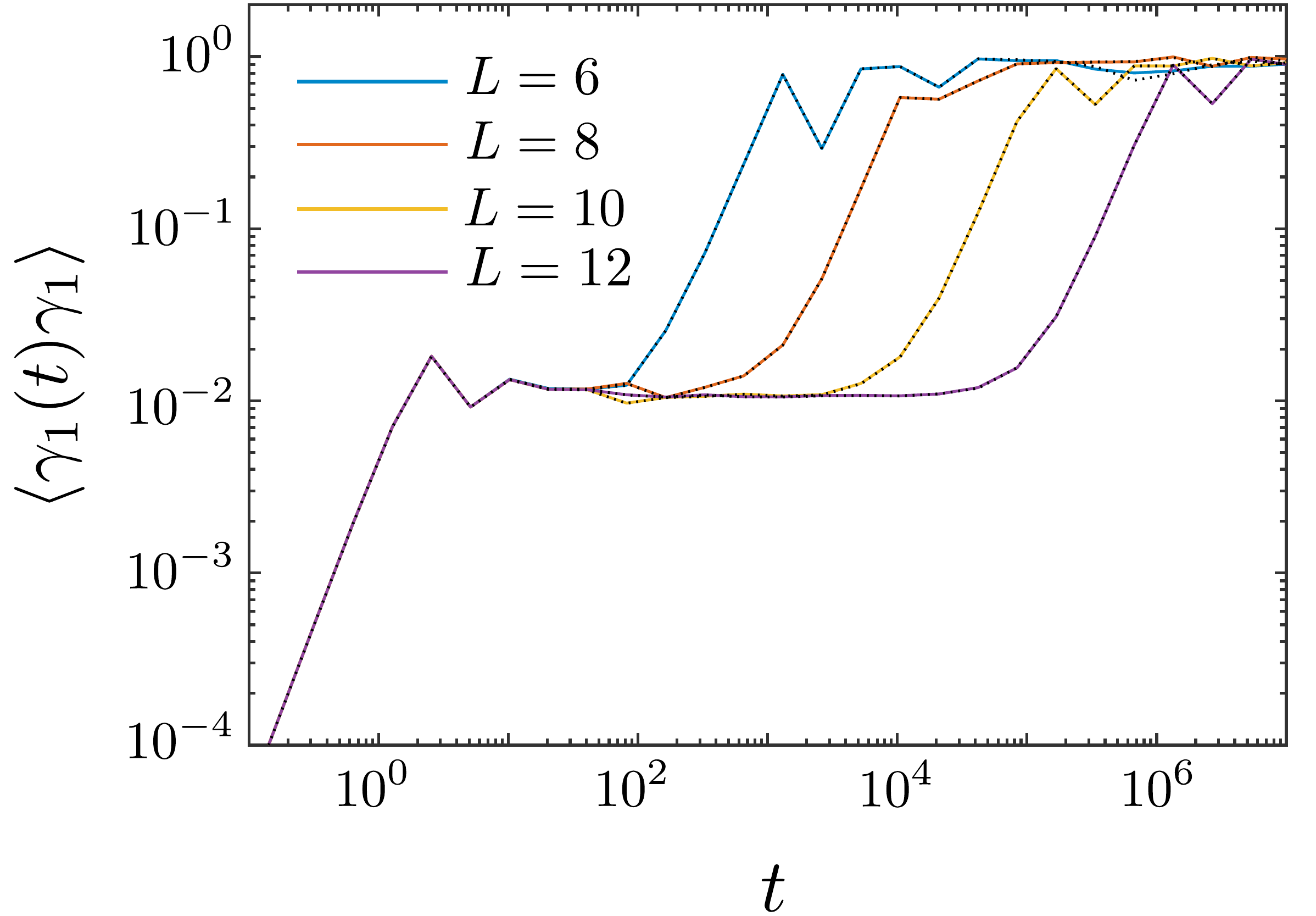}
\caption{Coherence of majorana fermion at the edge of the wire improves exponentially with increasing $L$ as expected in a Kitaev chain, but not with increasing $n_f$. Solid (dotted) lines correspond to $n_f = 1 (2)$.}
\label{fig:appfig1}
\end{figure}
\end{center}

\begin{center}
\begin{figure}[ht]
\includegraphics[width=3.5in]{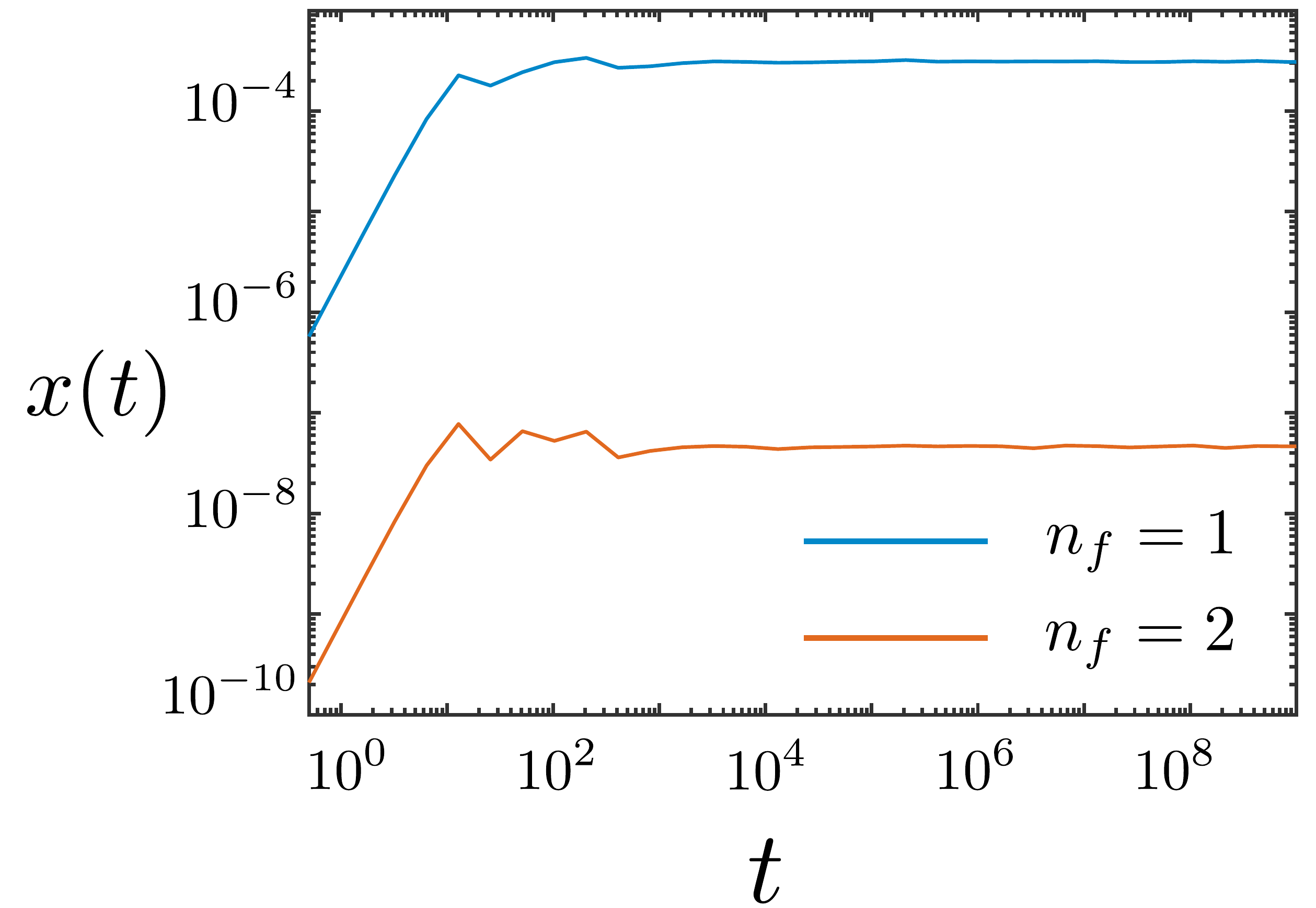}
\caption{Coherence of $P_X$, $x(t)$ improves with increasing $n_f$ in the same model as Fig.~\ref{fig:appfig2}.}
\label{fig:appfig2}
\end{figure}
\end{center}

The simplest demonstration of the above fact comes from numerical results of Figs.~\ref{fig:appfig1} and~\ref{fig:appfig2}. In the first figure, we examine the coherence of the majorana at one end of the system. There is initial relaxation that occurs on the microscopic time scales followed by a long plateau where coherence is maintained, before an eventual relaxation. As seen in Fig.~\ref{fig:appfig1}, the relaxation of the coherence occurs on a time scale that depends exponentially on length $L$ (equal linear displacements of the relaxation time scale can be seen on the log scale as $L$ is increased in fixed steps). In the same plot, dotted lines correspond to the protocol carried out at $n_f = 2$, while solid lines correspond to $n_f = 1$. In Fig.~\ref{fig:appfig2}, we plot the coherence $x(t)$ corresponding to the operator $P_X$. Clearly, this coherence is improved as $n_f$ is increased. It is important to recall that the protocol engineers a particular symmetry (in this case $P_X$) and there is an optimal number of fractal layers $n_f$ associated with it. Phases that rely on the preservation of said symmetry will be more robust the better this symmetry is implemented. In this particular instance, the creation of isolated majorana modes relies on the system being in the correct phase (with alternating weak odd and strong even bonds) and not on the engineering of additional global symmetries (besides the fermion parity which we assume to be conserved). Thus, the coherence of the majoranas only weakly depends on the fractal layers $n_f$. 

\section{Comments on Numerical Methods}

Time evolution has been performed using either exact diagonalization or recursive multiplication to get the unitary for time-evolution at exponentially long times. 

In the first method, we diagonalize $U(T_f)$. Subsequently, all operators are represented in the basis that diagonalizes $U(T_f)$. In this basis, the phase picked up by each individual Floquet eigenstate can be determined at time easily, and we do so at times that are uniformly spread out on a log scale. Coherences are evaluated as mentioned in the main text, averaging over a complete basis of initial states, except in the case of heating, where the initial state is the ground state of the Hamiltonian at zeroth order of the Magnus expansion, $H_{\vec{0}}$. This method usually works well except in instances where there is massive degeneracy in the spectrum; this makes precise basis rotation hard to compute when diagonalizing $U(T_f)$. In the example studied in the main text, this unfortunately limits us to studying time-evolution for $L = 4, 8, 12$ as $L = 6, 10, ...$ are degenerate, which further limits our ability to perform finite size scaling in a meaningul way with system sizes accessible. 

In the second method, which we use in instances where the spectrum has degeneracies (applicable to the Kitaev chain considered in the previous section), we simply compute unitary matrices for time-evolution at longer times by the recursion relation $U(2t) = U(t) \cdot U(t)$, for $t > T_f$. This resolves the issues with exact diagonalization in the presence of degeneracies, but is slower and suffers from the issue of the time-evolution matrix at late times becoming less unitary. This limits the dynamics to shorter times.

\end{document}